\documentclass[useAMS,usenatbib,rotating]{mn2e}
\usepackage{amssymb}     \usepackage{amsmath}    \usepackage{amsfonts}
\usepackage{xspace}    \usepackage{longtable}    \usepackage{rotating}
\usepackage{lscape}

\newcommand{\HI}{\mbox{\sc      H{i}}}

\newcommand{\mhi}{\mbox{$M_{\rm HI}$}}

\title[Misalignment between cold gas and stars in early-type galaxies]
  {Misalignment between cold gas and stellar components in early-type galaxies}
\author[O.\ I.\ Wong et al.]
  {O. Ivy Wong,$^{1}$\thanks{E-mail:ivy.wong@uwa.edu.au}  K.~Schawinski,$^{2}$ G.I.G.~J\'ozsa,$^{3,4,5}$ C.M.~Urry,$^{6,7}$ C.J.~Lintott,$^{8}$
  \newauthor B.D.~Simmons,$^8$ S.~Kaviraj,$^{8,9}$ and K.L.~Masters$^{10,11}$\\
  $^1$ International Centre for Radio Astronomy Research, The University of Western Australia M468, 35 Stirling Highway, Crawley,\\  \hspace{2mm} WA 6009, Australia\\
  $^2$ Institute for Astronomy, ETH Zürich, Wolfgang-Pauli-Strasse 27, 8093 Z\"urich, Switzerland\\
  $^3$ SKA South Africa, Radio Astronomy Research Group, 3rd Floor, The Park, Park Road, Pinelands, 7405, South Africa\\
  $^4$ Rhodes University, Department of Physics and Electronics, Rhodes Centre for Radio Astronomy Techniques \& Technologies,\\
 \hspace{2mm} P.O. Box 94, Grahamstown 6140, South Africa\\
  $^5$ Argelander Institut f\"ur Astronomie (AIfA), University of Bonn, Auf dem H\"ugel 71, 53121 Bonn, Germany\\
  $^6$ Yale Center for Astronomy and Astrophysics and Department of Physics, Yale University, P.O. Box 208120, New Haven,\\
 \hspace{2mm} CT 06520-8120, USA\\
  $^7$ Department of Astronomy, Yale University, P.O. Box 208101, New Haven, CT 06520-8101, USA\\
  $^8$ Oxford Astrophysics, Denys Wilkinson Building, Keble Road, Oxford OX1 3RH, UK\\
  $^9$ Centre for Astrophysics Research, University of Hertfordshire, College Lane, Hatfield, Herts, AL10 9AB, UK\\
$^{10}$ Institute of Cosmology \& Gravitation, University of Portsmouth, Dennis Sciama Building, Portsmouth, PO1 3FX, UK\\
$^{11}$ South East Physics Network (SEPNet), {\em{www.sepnet.ac.uk}}}
\date{Released 2011 Xxxxx XX}

\pagerange{\pageref{firstpage}--\pageref{lastpage}} \pubyear{2011}

\def\LaTeX{L\kern-.36em\raise.3ex\hbox{a}\kern-.15em
    T\kern-.1667em\lower.7ex\hbox{E}\kern-.125emX}

\begin{document}

\label{firstpage}

\maketitle

\begin{abstract}
Recent work suggests blue ellipticals form in mergers and migrate quickly from the blue
cloud of star-forming galaxies to the red sequence of passively evolving galaxies, perhaps
as a result of black hole feedback. Such rapid reddening of stellar populations implies
that large gas reservoirs in the pre-merger star-forming pair must be depleted on short time
scales. Here we present pilot observations of atomic hydrogen gas in four blue early-type galaxies
 that reveal increasing spatial offsets between the gas reservoirs and the stellar components of 
 the  galaxies,  with advancing post-starburst age. Emission line spectra show associated nuclear
 activity in two of the merged galaxies, and in one case radio lobes aligned with the
 displaced gas reservoir.  These early results suggest that a kinetic process (possibly feedback
from black hole activity) is driving the quick truncation of star formation in these systems, rather than a
 simple exhaustion of gas supply.

\end{abstract}


\section{Introduction}

\begin{figure*}[h]
\begin{center}
\includegraphics[scale=.6]{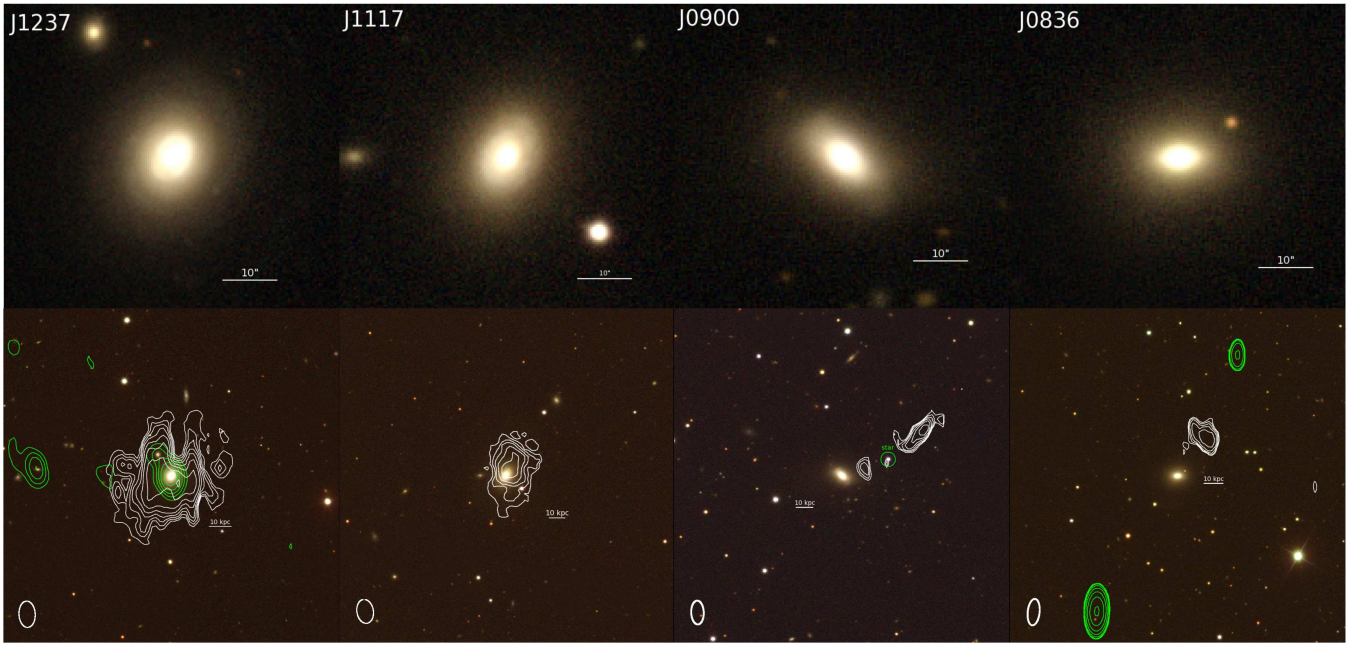} 
\end{center}
\caption{Top row: optical $gri$ SDSS \citep{ahn14} composite images of the four blue early-type galaxies in our sample. The galaxies are arranged from left to right in terms of $NUV-u$ colour from blue to red as a tracer of time. The white scale bars represent 10 arcseconds.  Bottom row: zoomed-out optical images with the WSRT radio observations overlaid. The white contours (asinh-stretch) represent the \HI\ and the green contours (logarithmic stretch) the radio continuum. The lowest radio and \HI\ contours begin at 3$\sigma$ and the highest contour is matched to the peak flux density of
0.87 mJy beam$^{-1}$, 1.51 mJy beam$^{-1}$, 0.86 mJy beam$^{-1}$, 44.79 mJy beam$^{-1}$ for J1237, J1117, J0900 and J0836; respectively.    It should be noted that the nuclear 1.4 GHz continuum point sources are not shown for J0900 and J1117 to avoid confusion with the \HI\ contours. 
}
\label{morph1}
\end{figure*}

Star-forming galaxies show a correlation between stellar mass and star formation 
rate whose normalisation varies with redshift, but whose scatter remains tight 
out to high redshift \citep{noeske07,peng10,elbaz11}. This correlation can be explained
 as a balance between gas inflows from cosmological scales and outflows driven by 
supernovae \citep{bouche10,lilly13}. Quenching of star formation causes galaxies to
 depart from this steady-state along pathways that depend on the evolution of the gas 
supply and reservoir \citep{schawinski14}: if the cosmological gas supply to a galaxy 
is shut off, it will slowly exhaust its gas reservoir by forming stars at a declining
 rate. If instead the gas reservoir fuelling star formation is destroyed effectively
instantaneously, then star formation will cease and the galaxy will redden rapidly
 within ∼ 1 Gyr, as is observed in early-type galaxies \citep[ETGs; ][]{kaviraj11,wong12,schawinski14}.
Interestingly, this timescale of 1 Gyr is consistent with that proposed for the
 transition of a merger-driven Ultraluminous Infrared Galaxy (ULIRG) 
to an ETG \citep{emonts06}. Also, it has been known for a while that ULIRGs have a 
high merger fraction \citep{sanders88}.

In addition, negative feedback from a galaxy's central active galactic nuclei has 
often been invoked in galaxy formation models to slow down the star formation history 
of simulated galaxies \citep[e.g.\ ][]{croton06}.  However, observational evidence
for such feedback is currently only available for a few individual galaxies 
\citep{harrison14,nyland13,alatalo11,hota11,croston08,kharb06}.  With the advent of
very large surveys;  much progress has been made towards  understanding of the 
co-evolution between galaxies and their central AGN, in particular, the connection between
classical bulges and central supermassive black holes  \citep{heckman14,kormendy13}.

Blue ETGs do not fit the canonical bimodal scheme of red ellipticals versus blue spirals
and are unlikely to be the descendants of blue spirals \citep{tojeiro13}.  Rather, the 
blue ETGs appear to be transition-type galaxies and are
 probable predecessors of local post-starburst galaxies \citep{schawinski09,wong12}.
  Previous CO observations of blue ETGs show that the molecular gas reservoirs are being rapidly 
destroyed during the composite star formation and AGN phase \citep{schawinski09a}.
Local post-starburst galaxies are defined to be galaxies which have recently ceased 
forming stars.  These galaxies typically 
show little H$\alpha$ or [OII] emission (indicative of current star formation), 
but have strong absorption line signatures indicative of a young stellar 
population.  In addition, local post-starbursts consists mostly of galaxies
which have similar early-type structural properties to red sequence galaxies of
comparable masses \citep{wong12}.


Earlier attempts to map the \HI\ content of post-starburst galaxies 
 only found \HI\ in  one of five targets
 which consisted of an interacting pair of galaxies \citep{chang01}.
  All subsequent campaigns to detect \HI\  in post-starbursts have been optimised
 towards detecting diffused low-surface brightness gas, while compromising on
the angular resolution required to map the location of the gas \citep[e.g.\ ][]{zwaan13}.
  
To determine the physical process responsible for shutting down star formation,
we posit that post-starburst galaxies are too evolved and that {\em{the smoking gun 
for quenching is more likely to be found in its predecessor population, namely, 
the blue early-type galaxies.}}

Using the Westerbork Syntheses Radio Telescope, we study the \HI\ content of a pilot 
sample of four blue early-type galaxies---the progenitors to post-starburst 
galaxies because 1) these galaxies  are still currently star-forming and therefore
 there should be gas where there are stars being formed;  and 2) the state of the 
gas morphology and dynamics will shed light on why these galaxies will soon stop
 forming stars.




Section 2 describes the blue early-type sample.  The pilot observations and data processing 
methods are described in Section 3. We discuss the radio continuum and \HI\ imaging
 results in Section 4 and Section 5, respectively.  Section 6 provides a summary of our results.
The AB magnitude system is used throughout this work.

\section{Blue early-type galaxies}

We obtain the photometric and spectroscopic data from the Sloan  Digital Sky Survey (SDSS)
 DR7 \citep{york00,abazajian09} for all objects classified as `galaxy' \citep{strauss02}. 
The sample of 204 low-redshift ($0.02 < z < 0.05$) blue ETGs were identified using the Sloan
 Digital Sky Survey \citep[SDSS; ]{adelman08} and the Galaxy Zoo project 
\citep{lintott08,schawinski09,lintott11}.  The blue ETGs account for 5.7\% of all ETGs found
 within the same redshift range and are the most actively star-forming population of ETGs
 \citep{schawinski09}. Further description of the blue early-type galaxy selection can be 
found in \citet{schawinski09} and at {\tt{http://data.galaxyzoo.org}}.  

\subsection{The pilot sample}
\begin{figure}
\begin{center}
\includegraphics[scale=.37,angle=90]{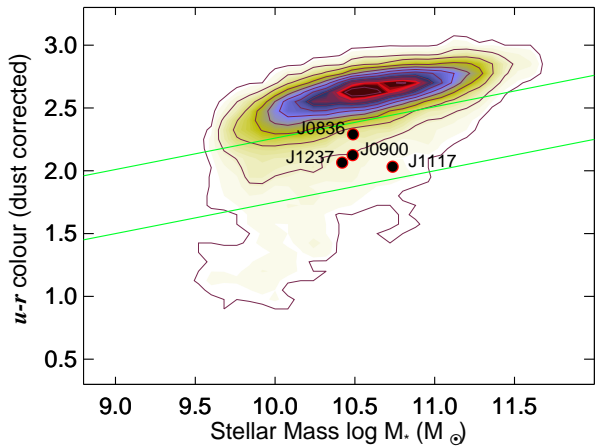} 
\end{center}
\caption{Colour-stellar mass diagram of ETGs found from SDSS and Galaxy Zoo. The contours
represent the number density of ETGs occupying the colour-stellar mass region known as 
the `red sequence'.   All four pilot sample galaxies
are located in the optical `green valley' of the colour-mass diagram (as demarcated between the solid 
green lines). All galaxies show the presence of intermediate age stellar
populations by their `green' $u-r$ colour.}
\label{colmass}
\end{figure}

To investigate the fate of the gas reservoir when star formation is quenched,
 we select four ETGs with comparable stellar masses and 
UV/optical colours indicative of rapid quenching: all four lie in the optical 
$u-r$ green valley, indicating they have significant intermediate-age stellar 
populations, and their $NUV-u$ UV/optical colours range from very blue,
consistent with ongoing star formation (J1237+39), to the redder colours of 
passive, quenched galaxies (J0836+30)---see Figure~\ref{colmass}. Due to the 
diffused nature of \HI\ gas, we have also selected the nearest galaxies which can be 
observed by the Westerbork Synthesis Radio Telescope (WSRT).  
The 
optical/stellar morphologies of this sample are presented in the top row of Figure~\ref{morph1}. 


\begin{figure}
\begin{center}
\includegraphics[scale=.37,angle=90]{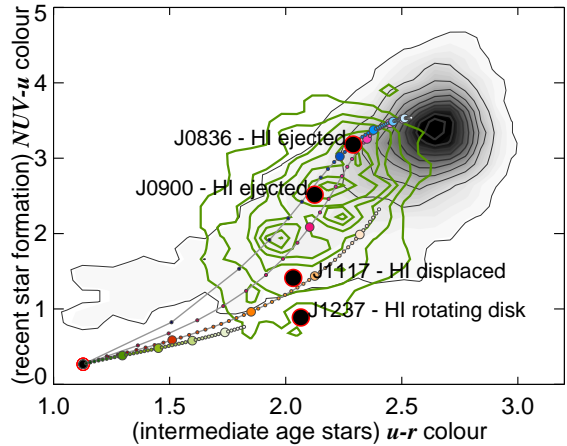} 
\end{center}
\caption{The UV/optical colour-colour diagrams of blue early-type galaxies (green density contours)
and the entire early-type galaxy population (black shaded contours). The model stellar population tracks
with varying quenching timescales from \citet{schawinski14} are also overlaid.  The large and small
 circles along the evolutionary tracks represent 1 Gyr and 100 Myr, respectively.
Two out of four pilot sample galaxies show an absence of very young stellar
populations indicated by red $NUV-u$ colours. Those three galaxies also show either a displaced
HI reservoir (J1117+51) or a completely ejected HI reservoir (J0836+30 and J0900+46). 
The galaxy which is bluest in $NUV-u$ features a rotating HI disk.}
\label{colcol}
\end{figure}

Stellar population modelling find two distinct star formation truncation timescales for early- and 
late-type galaxies whereby  the early-type galaxies appear to have fast quenching timescales ($\leq$1 Gyr), 
while the late-type galaxies have quenching timescales of several Gyrs  \citep{schawinski14}.
Figure~\ref{colcol} illustrates the different evolutionary stages represented by each
of the four ETGs selected for this pilot study on a UV/optical colour-colour diagram.
The colour of model stellar populations along different star formation quenching tracks 
are shown as solid lines where each large circle marks 1 Gyr and intervals of 100 Myr being shown by small circles along
the stellar evolution tracks.  The black shaded contours represent the low-redshift
early-type population and the green contour represents the population of blue
early-type galaxies selected from \citet{schawinski09}. The observed stellar
populations of J0836+30 and J0900+46 appear to be relatively evolved along the fastest 
quenching evolutionary pathway that occurs on timescales of several hundred 
Myr to 1 Gyr \citep{schawinski14}.  On the other hand, the stellar population colours of J1117+51 and J1237+39
are less evolved and consistent with earlier stages of quenching.  We note that J1117+51 and J1237+39 appear
to favour slower quenching pathways in Figure~\ref{colcol}. However, this could be due to differences
between the start times for quenching in these galaxies relative to the fiducial starting point for quenching in the models. 
The \HI\ properties of each sampled galaxy appears to be consistent with 
its respective evolutionary stage.  See Section 5 for more details of the observed \HI\ properties
of this sample.

\begin{figure*}
\begin{center}
\includegraphics[scale=.75]{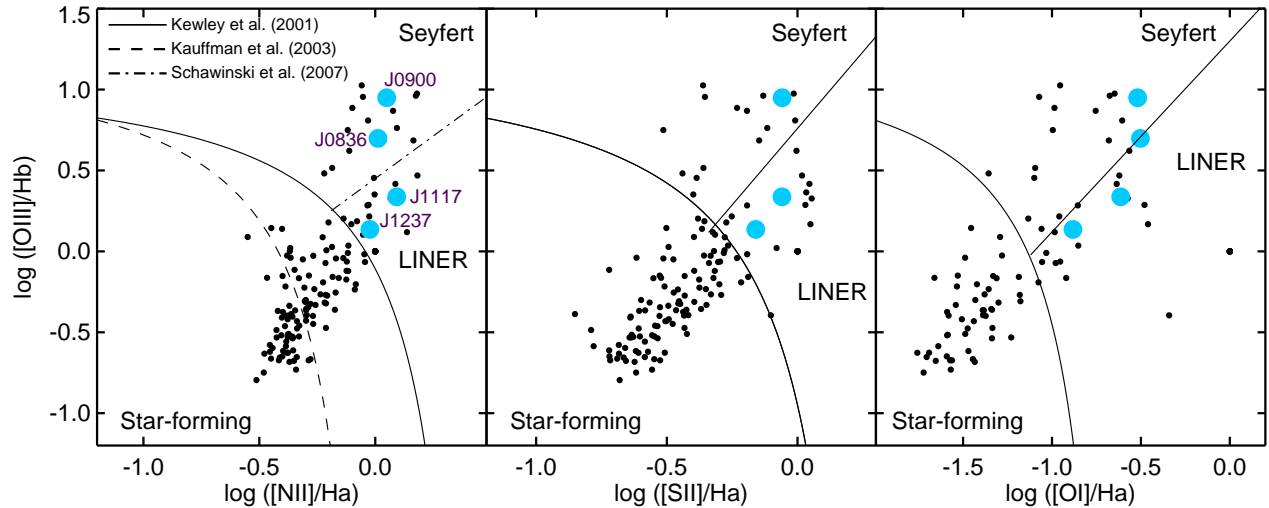} 
\end{center}
\caption{Optical nebular emission line diagrams. The nebular emission line
 ratios of the entire Schawinski et al. (2009) sample of blue early-type galaxies
 are plotted as black points.  The blue solid points represent the objects selected in
 this pilot sample. The solid and dashed line are two lines used to separate the nebular
emission originating from star formation rather than that of Seyfert or LINER activity
\citep{kewley01,kauffmann03}. The dotted-dashed line differentiates between ionisation
levels obtained from LINER versus Seyfert cores \citep{schawinski07}.}
\label{bpt}
\end{figure*}

In addition, all four galaxies show optical emission lines dominated by activity
 other than star formation: J0836 and J0900 show Seyfert-like lines while 
J1117 and J1237 show low-ionization lines which could be due to black hole 
accretion, shocks or evolved stars \citep{sarzi10}. Figure~\ref{bpt} 
 presents the Baldwin, Phillips \& Televich (BPT) diagnostic diagram 
\citep{baldwin81} which shows the emission line ratios observed for 
blue early-type galaxies. Table~\ref{genprop} lists the optical 
properties of this pilot sample.  Stellar masses were obtained from 
spectral fitting \citep{schawinski09}.

\begin{table*}
\footnotesize{
\begin{center}
\caption{Optical properties of our low-redshift blue early-type galaxy sample.}
\label{genprop}
\begin{tabular}{llccccccc}
\hline 
\hline
SDSS ID & Galaxy & RA (J2000) &Declination (J2000) & Redshift & Distance & $r$ &$u-r$ & log($M_{\star}$) \\ 
(1) & (2) & (3) & (4) &(5) & (6) &(7) & (8)& (9)\\
\hline 
587735044686348347& J0836+30  & 08:36:01.5 &  +30:15:59.1& 0.02561 &105 &-21.1 &  2.27 & 10.7 \\ 
587731887888990283 &J0900+46 & 09:00:36.1 & +46:41:11.4 & 0.02748 &113 &-21.3 &  2.20 & 10.5 \\
587732135382089788 &J1117+51 & 11:17:33.3 & +51:16:17.7 &  0.02767 & 115 &-21.4&  2.33 & 10.6 \\
587738946685304841& J1237+39 & 12:37:15.7 & +39:28:59.3 &  0.02035 & 84 &-20.9&  2.16 & 10.3 \\

\hline
\hline
\end{tabular}
\end{center}}
Col.\ (1): SDSS object identification.  Col.\ (2): Galaxy identification used in this paper. Col.\ (3): Galaxy center's right ascension.  Col.\ (4): Galaxy center's declination. Col.\ (5): Redshift. Col.\ (6): Distance in Mpc. Col.\ (7): SDSS $r$-band magnitude. Col.\ (8): Optical $u-r$ colour. Col.\ (9): log (stellar mass) in solar mass as estimated by spectral fitting.
\end{table*} 

\section{Pilot survey}
\subsection{Radio observations}
We performed imaging of the 1.4 GHz radio continuum and the atomic 
Hydrogen (\HI) emission of our pilot sample using the WSRT in the Netherlands between June and November 2012. 
As the WSRT is an east-west interferometer, our observations were divided 
into several epochs to optimise the $uv$-coverage. The single 20 MHz band 
is sampled over 1024 channels in the maxi-short baseline configuration in order
to obtain a spectral resolution of 4 km s$^{-1}$ spanning the range of 
 4000 km s$^{-1}$.
\begin{figure}
\begin{center}
\includegraphics[scale=.31]{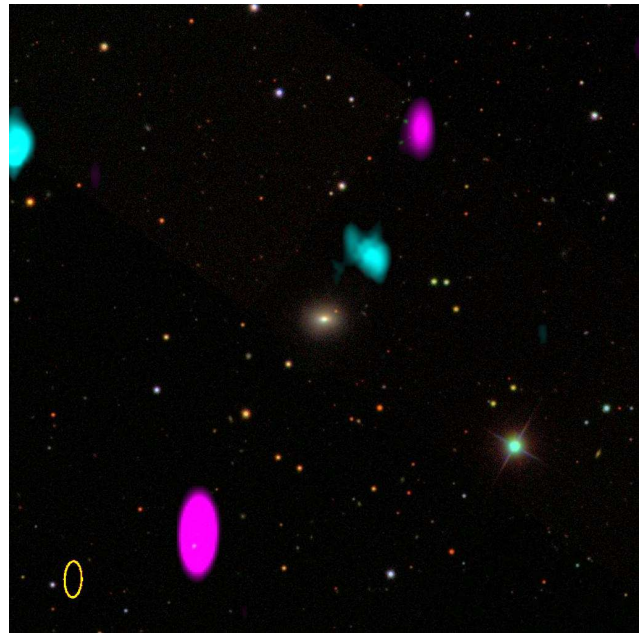} 
\end{center}
\caption{Multicolour composite of the J0836+30 field (centred on J0836+30) 
where the background colour image is a multicolour composite of the 
five SDSS optical $ugriz$ bands.  The 1.4 GHz radio continuum and 
\HI\ emission from these pilot observations  are shown in 
magenta and cyan, respectively. The yellow ring in the bottom-left of the
figure represents the beam of the radio observations. }
\label{multicol}
\end{figure}

Using this setup, a 24-hour on-source integration should result in an 
RMS noise level of 0.54 mJy beam$^{-1}$ at full angular resolution over a  
FWHM of 8.25 km s$^{-1}$ in velocity---corresponding to a column density
sensitivity of $1.46\times 10^{19}$ atoms cm$^{-2}$ per resolution assuming
a uniform weighting\footnote{Noise estimations are obtained 
from the WSRT exposure time calculator found at 
{\tt{http://www.astron.nl/~oosterlo/expCalc.html}}}. Due to timetabling constraints, 
we were not able to obtain 24-hour on-source integration for each 
of the four galaxies. Total integration times for each 
 galaxy, the resultant synthesised beam properties and sensitivities
 are listed in Table~\ref{obs}. 

\begin{table*}
\footnotesize{
\begin{center}
\caption{WSRT observation summary. WSRT observation integration times and resulting beam properties are detailed for each of our pilot blue galaxies.}
\label{obs}
\begin{tabular}{llccccc}
\hline 
\hline
Galaxy & Observations dates & Total integration time & Beamsize & Beam Position angle & RMS$_{\HI }$ & RMS$_{1.4}$\\ 
(1) & (2) & (3) & (4) &(5) &(6) &(7)\\
\hline 
J0836+30 & Nov 15,25 & 20.3 &29.5, 13.9  &  -2.0 & 4.1 & 0.083\\
J0900+46 & Oct 23, Nov 11 & 10.3 & 24.0, 12.1  &  -1.7 & 11.0 & 0.062\\
J1117+51 & June 30, Nov 18 & 24.0 &19.8, 14.5  &   9.6 & 3.7 & 0.082\\
J1237+39 & Aug 24, Oct 25, Nov 21 & 24.2 &27.6, 17.5  &   4.7 & 32.6 & 0.038\\
\hline
\hline
\end{tabular}
\end{center}}
Col.\ (1): Galaxy identification.  Col.\ (2): Observation dates. Col.\ (3): Total on-source integration time in hours.  Col.\ (4): Synthesised beam diameter in arcseconds.  Col.\ (5): Synthesised beam position angle in degrees.  Col.\ (6): RMS noise level for the \HI\ emission in mJy beam$^{-1}$.  Col.\ (7): RMS noise level for the 1.4 GHz radio continuum in mJy beam$^{-1}$.
\end{table*} 

\subsection{Data processing}
The observations are calibrated and processed in the standard
manner using the {\sc{Miriad}} Reduction software \citep{sault95}.  
After the initial bandpass calibration using a WSRT standard calibrator, 
continuum images are produced and improved by iteratively applying a 
continuum self-calibration. The resulting calibration tables are then 
used for the line imaging as well. After a second continuum subtraction, 
the resulting \HI\ image cubes are produced by using a robust weighting 
of 0.4 in order to produce the optimal balance between angular resolution 
and surface brightness sensitivity. We average across every second 
channel to improve the signal-to-noise sensitivity  which results in 
a velocity resolution of $~$8.5 kms$^{-1}$ (after Hanning smoothing).  
It should be noted that these final reduction parameters  were 
determined after we had already tested various combinations of weighting schemes (including 
natural and uniform weightings) with different angular and frequency tapering
for each set of observations.
Table~\ref{obs} lists the RMS levels obtained for each galaxy 
in our sample.

\section{Radio continuum properties of blue early-type galaxies}

We observe 1.4 GHz radio continuum emission in the nuclear regions of
 J0900+46, J1117+51 and J1237+39.  No nuclear radio emission was found 
for J0836+30,  but we do observe two radio lobes extending 
88.4 kpc north-west and 102.5 kpc south-east of J0836+30 along 
the same direction as the extragalactic \HI\ cloud.  
Figure~\ref{multicol} shows a multicolour composite image of the 
J0836+30 field where the radio continuum and the integrated \HI\
 emission is shown in magenta and cyan, while the background
three-colour image is produced from the five SDSS $ugriz$ bands.
Further discussion of the possible mechanism for the displacement 
of the \HI\ reservoir in J0836+30 can be found in Section 5.1.

We argue that the two radio lobes flanking both sides of J0836+30 are likely 
to be faded radio lobes belonging to J0836+30 and not likely to be due to a 
chance alignment with background sources.  Although there is some overlap 
between the South-Eastern radio lobe with a WISE 3.6 micron source, we do not
find an optical/IR counterpart for the North-Western radio lobe.
In addition, the projected separation between both lobes and J0836+30 are 
similar enough for the differences in angular separation (as well as brightness) 
to be accounted for by projection effects. 


The nuclear radio luminosities  range from  $2.0 \times 10^{35}$ to $7.4 \times 10^{35}$ ergs s$^{-1}$
(or $1.0 \times 10^{20}$ to $3.7 \times 10^{21}$ W Hz$^{-1}$) and  are consistent 
with those found mostly  in other radio studies of low-redshift
 star-forming galaxies \citep{mauch07,jarvis10}. For  J0836+30, 
the luminosities of the radio lobes in the north-west and south-east
directions are $9.26 \times 10^{35}$ and $1.22 \times 10^{37}$ ergs s$^{-1}$, respectively.
 Table~\ref{radprop} lists the measured 1.4 GHz radio continuum properties.
Figure~\ref{morph1} shows the radio continuum emission as green
contours and the \HI\ emission as white contours for the entire pilot sample.

Assuming that the nuclear radio continuum emission is due to star formation,
we can estimate  the star formation rate upper limits in our pilot
sample.  Using the star formation rate calibrations from 
\citet{bell03,hopkins03}, we find that our galaxies have star 
formation rates that are fewer than 2 solar mass per year. 
J1117+51 has the highest estimated star formation rate with
 1.7 solar mass per year.

Archival IRAS 60 $\mu$m observations \citep{moshir90} were found for 
J1117+51 and J1237+39.  Using the far-infrared to radio correlation (FRC) of 
low-redshift star-forming galaxies \cite{yun01},  we find that the expected 
far-infrared 60 $\mu$m luminosity that correlates to our measured 
radio luminosity if the radio emission is due solely to star formation
is log $L_{60}$ of 9.6 and 9.0 for J1117+51 and J1237+39, respectively.
In comparison to the IRAS log $L_{60}$ values of 9.4 and 9.3, we find that
the predicted values from the FRC are consistent with the idea
that the 1.4 GHz continuum emission originates from star
formation for both J1117+51 and J1237+39.  It should be noted that the
FRC has a scatter of approximately 0.25 dex and the IRAS measurement 
uncertainties are at the 20\% level.

\begin{figure*}
\begin{center}
\includegraphics[scale=.7,angle=0]{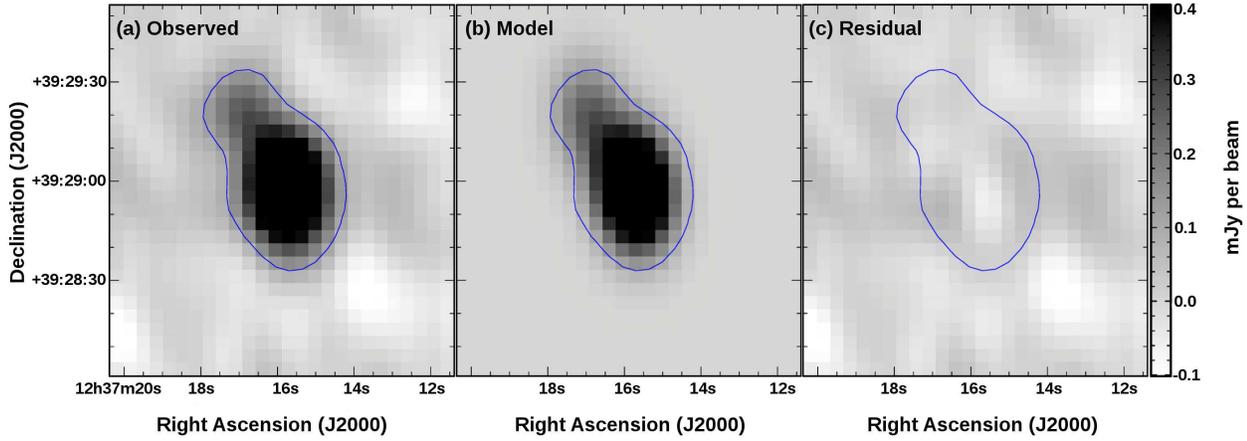}
\end{center}
\caption{The 1.4 GHz radio continuum emission from J1237+39. Centred
on J1237+39, we model the emission as two overlapping point sources. 
The solid blue line in all three panels mark the $3\sigma$ flux level 
of the observed 1.4 GHz emission (see Panel a). Panels b and c show
the model of two point sources and the residual maps, respectively.  }
\label{imfit}
\end{figure*}

Although the radio continuum emission observed from J1237+39 appears to be extended,
it is in fact the result of a chance alignment between J1237+39 and a projected background galaxy, 
SDSS J123716.92+392921.9,  which is 8.5 times more distant 
(at redshift $z=0.176$) than J1237+39.  Figure~\ref{imfit} shows that
 the observed radio emission can be simply modelled by two point sources.  
The lack of any regions which display a flux excess or a flux deficit 
in our residual image (panel c of Figure~\ref{imfit})  is consistent 
with the idea that the radio emission from J1237+39 does not have an extended 
diffuse component.

\begin{table*}
\footnotesize{
\begin{center}
\caption{Radio (1.4 GHz) and \HI\ properties of blue early-type galaxies.}
\label{radprop}
\begin{tabular}{lcccccccc}
\hline 
\hline
Galaxy & $S_{1.4}$ & $L_{1.4}$ & $L_{1.4}$ & SFR$_{1.4}$ & $S_{\HI}$ & $v_{\HI}$ & $w_{\HI}$ &\mhi\ \\
(1) & (2) & (3) & (4) &(5) &(6) &(7) & (8)& (9)\\
\hline 
J0836+30 & $<2.5 \times 10^{-4}$ & $<3.3 \times 10^{20}$ $^{**}$& $<6.6 \times 10^{34}$ & $<0.4$ & 0.078 (0.025) & 7705 & 140 & 2.0 (0.7)\\
J0900+46 & $9.7 (2.4) \times 10^{-4}$& $1.5 \times 10^{21}$ &$3.0 \times 10^{35}$& 1.1 & 0.410 (0.182) & 8277 &  62 & 12.3 (5.5) \\
J1117+51 & $23.4 (3.8) \times 10^{-4}$& $3.7 \times 10^{21}$&$7.4 \times 10^{35}$  & 1.7 &0.100 (0.020) & 8185 & 140 & 3.1 (0.6) \\
J1237+39 & $12.0 (1.6) \times 10^{-4}$& $1.0 \times 10^{21}$&$2.0 \times 10^{35}$  & 0.7 & 2.448 (0.403) & 6096 & 245 & 40.8 (6.7)\\
\hline
\hline
\end{tabular}
\end{center}}
$^{**}$Note that the luminosities of the radio lobes in the north-west and south-east directions are
$4.63\; (\pm0.44)\; \times 10^{21}$ WHz$^{-1}$ and $6.12\; (\pm0.03)\; \times 10^{22}$ WHz$^{-1}$, respectively. 
 Col.\ (1): Galaxy identification.  Col.\ (2): Total 1.4 GHz radio continuum emission in Jansky. Col.\ (3): Radio continuum luminosity (1.4 GHz) in WHz$^{-1}$.  Col.\ (4): Radio continuum luminosity (1.4 GHz) in ergs s$^{-1}$.   Col.\ (4): Estimated star formation rate using the calibration by \cite{bell03,hopkins03} in M$_{\odot}$ year$^{-1}$.  Col.\ (5): Integrated \HI\ emission in Jansky.  Col.\ (6): \HI\ radial velocity as measured from the mid-point at full-width half-maximum in kms$^{-1}$.  Col.\ (7): The 50\%  \HI\ velocity width in kms$^{-1}$. Col.\ (8): Total \HI\ mass in units of $\times 10^8$M$_{\odot}$.  Values in brackets give the estimated errors. 
\end{table*}

\section{\HI\ content of blue early-type galaxies}

\begin{figure*}
\begin{center}
\includegraphics[scale=1.,angle=0]{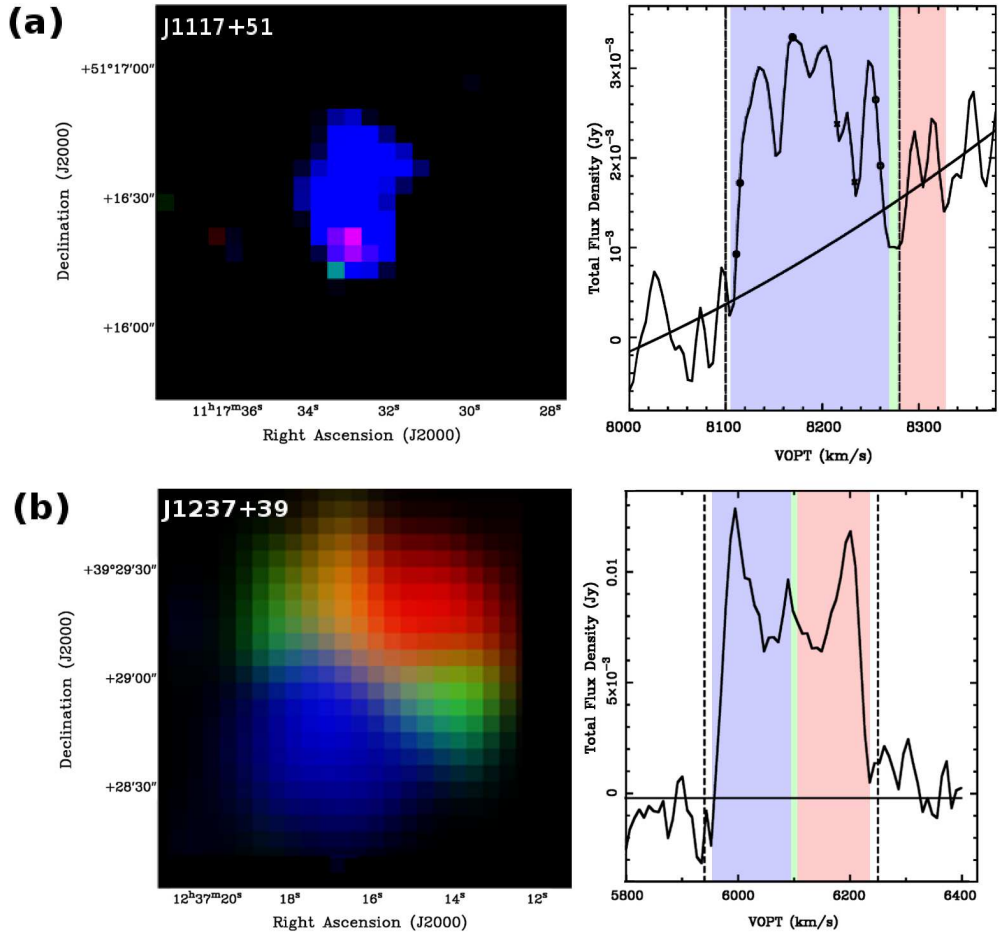}
\end{center}
\caption{Velocity maps and spectra of J1117+51 (panel a) and J1237+39 (panel b). 
  The 3-color velocity maps correspond to the \HI\ emission in the velocity 
range highlighted in the same color on the corresponding velocity spectrum.  
It should be noted that the green 
represents the optical velocity which we assume to be the systemic velocity of
each of the blue early-types sampled in this paper.  }
\label{velspec2}
\end{figure*}

As the four blue ETGs in our sample represent four separate stages of 
evolution (as traced by the UV/optical colours and the [OIII]/[H$\beta$] ratios),
 we observe a correlation with their respective \HI\ gas-to-stellar mass fractions
 whereby the gas fractions decrease as a function of time with increasing UV/optical colours
and increasing [OIII]/[H$\beta$] ratios.  Table~\ref{radprop} lists the measured
\HI\ properties.  The galaxy at the earliest stage of quenching in our sample (i.e.\  the galaxy with the 
bluest NUV/optical colour and the weakest [OIII]/[H$\beta$] ratio), J1237+39,  features
a central, undisturbed, rotating HI disk.  Both spatial and kinematic 
asymmetries are not observed in this galaxy (see Figure\ref{morph1} and 
Figure~\ref{velspec2}).

In the second bluest galaxy, J1117+51,  the \HI\ gas 
appears to be mostly  within J1117+51. However the gas morphology is 
fairly asymmetric and appears offset from the optical centre of the galaxy.
Assuming the optical velocity to be the systemic velocity of each galaxy, 
we find the \HI\ gas in J1117+51 to be largely blue-shifted by up to 170 kms$^{-1}$ 
from the galaxy's systemic velocity (see Figure~\ref{velspec2}).

\begin{figure*}
\begin{center}
\includegraphics[scale=1.,angle=0]{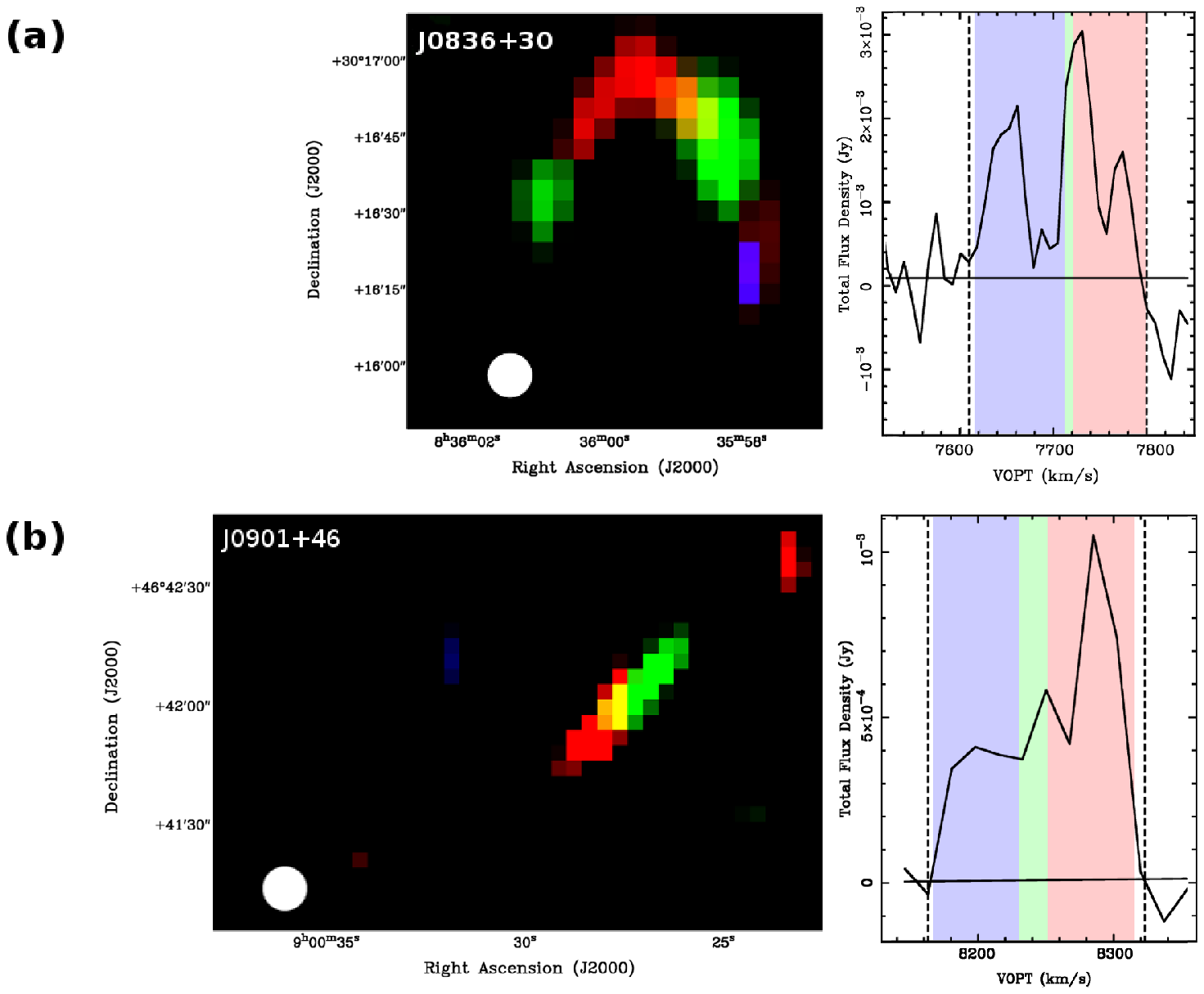}
\end{center}
\caption{Velocity maps and spectra of J0836+30 (panel a) and J0900+46 (panel b). 
  The 3-color velocity maps correspond to the \HI\ emission in the velocity 
range highlighted in the same color on the corresponding velocity spectrum. 
The white solid circle represents the position of the blue early-type galaxy
 relative to the observed \HI\ emission. It should be noted that the green 
represents the optical velocity which we assume to be the systemic velocity of
each of the blue early-types sampled in this paper. }
\label{velspec1}
\end{figure*}

In the redder galaxies and the galaxies with stronger [OIII]/[H$\beta$] ratios---indicative 
of ionisation levels consistent with Seyfert activity, the \HI\ reservoirs of J0836+30
and J0900+46 are completely displaced from their respective galaxy centres by 14--86 kiloparsecs.
In addition,  we find a significant amount of red-shifted \HI\ (by $\approx 70-80$ kms$^{-1}$)
 with respect to the systemic velocity of the individual galaxy (see Figure~\ref{morph1}).  

In J0836+30, the entire gas reservoir is offset spatially by 1 arcminute (projected distance of 30.5 kpc).
To approximate the timescale for the displacement of the gas reservoirs from J0836+30 and J0900+36,
 we simply divided the projected distances of the gas clouds by the observed \HI\ velocity offsets from
the systemic velocities taken from the optical observations.  For J0836+30, the  
timescale for the gas reservoir to have reached its current projected distance is approximately
373 Myr.  
Similarly, the \HI\ emission detected near J0900+46 spans between
14 kpc to 86 kpc from the optical center of J0900+46 and appears to be redshifted by 
approximately 70 kms$^{-1}$ (Figure~\ref{velspec1}b).  We estimate the displacement timescale to range 
between 0.2 to 1.2 Gyr.

\subsection{Mechanisms for the removal of \HI}

Support for the hypothesis that the observed \HI\ clouds in the vicinity 
of J0836+30 and J0900+46  have been removed from their parent galaxy
come from two main arguments.
 Firstly, the proximity of the gas clouds are very similar 
to the distances between the Milky Way and its high velocity clouds (HVCs).
Secondly, neither J0836+30 nor J0900+46  reside in dense environments with close neighbouring
galaxies which could be responsible for stripping the gas reservoirs from each of these galaxies.
In fact, \citet{verley07} has that found both J0836+30 and J0900+46 to be isolated galaxies
using different galaxy isolation metrics.   Several other previous studies have also classified
J0836+30 to be a prototypical isolated galaxy \citep{verdes05,stocke04,karachentseva73}.

In dense galaxy environments, such as clusters or compact galaxy groups,
intergalactic gas clouds are not uncommon \citep{kilborn00,oosterloo05,borthakur10}
 due to various gravitational 
and hydrodynamic interaction processes such as  galaxy-galaxy
interactions \citep{mihos04}, harassment \citep{moore96,moore98},
ram pressure and turbulent viscous stripping \citep{vollmer01}.

Ram pressure stripping is unable to shift such large amounts of gas
 using simple Toomre \& Toomre pressure arguments \citep[e.g.\ ][]{chung07}.  
Also, the galaxies in this pilot sample do not have nearby galaxy neighbours
 nor do they reside in dense galaxy environment such as clusters or compact groups.

Nearby examples where the majority of the galaxy's gas is displaced via tidal 
stripping also results in strong morphological distortions (e.g.\ NGC 4438 in
 the Virgo Cluster \citep{hota07}).  \citet{hota12} has also identified
a post-merger system, NGC 3801, where the extraplanar \HI\ resulting from the merger
has yet to be encountered by a newly-triggered radio jet.  As we do not observe any strong optical
 signatures from our galaxies such as stellar tidal tails, we can only infer
that (1)the tidal features resulting from the strong interaction that removed the
 large gas mass have faded; or (2)another physical mechanism is responsible for 
the gas removal.  

In the latter case, we posit that an active central engine has the required energy
 to expel a galaxy's gas.  In the case of J0836+30, we observe two radio lobes in
 the same direction as the ejected \HI\ cloud.  This is suggestive that at earlier
epochs, J0836+30 hosted a radio AGN which may have blown out 
the gas cloud that we observe. This hypothesis is consistent with recent studies
which have found nuclear activity to be more prevalent in star-forming galaxies relative to weakly star-forming or quenched galaxies \citep{rosario13}.

 As  the displacement timescale for this cloud is estimated
to be $\approx 373$ Myr, it is conceivable that J0836+30 has since evolved to a more
Seyfert state (as observed via the optical nebular emission).  A test of 
this hypothesis would be the presence of ionized gas adjacent to and between
 the observed \HI\ cloud and the galaxy. 
Therefore, in the case of J0836+30, we think that the alignment between
the radio lobes and the extragalactic \HI\ cloud is more suggestive of 
gas ejection via radio jets than a previous tidal encounter whose tidal signatures
have since faded in the past 373 Myr.

Previous observations of AGN-driven gas outflows 
have so far been from  studies where the outflows are determined from the 
kinematic structures of the emitting or absorbing gas \citep{gopal00,nesvadba09,mahony13,morganti13}.
However, recent observations of $^{12}$CO(1-0) emission from high-redshift radio galaxies
have found spatial offsets between the molecular gas reservoir and their host galaxies 
on the scales of several tens of kiloparsecs and aligned in the same direction to the radio
hotspots/lobes \citep{emonts14}.  We propose that we are witnessing a similar ejection of 
cold gas in J0836+30.

In the case of J0900+46, we neither observe strong radio continuum emission
nor nearby galaxies which could remove the gas from this galaxy.

We posit that the observed extragalactic gas clouds in J0836+30 and J0900+46 are
 outflows and not inflows for two reasons. Firstly, if the observed cold clouds were 
being accreted, the source of this gas is either a neighbouring galaxy or primodial 
gas leftover from galaxy formation. However, neither galaxies have neighbouring 
galaxies and the accretion of primodial gas at the current epoch is in a warm-hot phase
 rather than that of the observed cold \HI.   Secondly, \citet{bouche12} found that
 gas outflows are usually found at angles nearly orthogonal ($>60$ degrees) to the plane of the galaxy, 
whereas inflows have small azimuthal angles ($<30$ degrees) the galactic plane.
Therefore, we think that it's more likely that our observed extragalactic \HI\ clouds
originated from our target galaxies.

The mass of our gas clumps are approximately the total mass of HVCs around 
the Milky Way, the LMC and the SMC. 
Assuming that our observed gas clumps are analogous to these HVCs, we can expect
a similar gas dissipation time on the order of several hundred million years 
 if there isn't some sort of support mechanism that prevents  or slow down 
against cloud dissipation \citep{putman12}. The cloud 
survival time is linked closely to its total mass, cloud density, relative
 halo density and velocity \citep{putman12}. Previous studies 
have found that while the least bound material is likely to expand out
 into the IGM, bound structures are likely to fall back onto the galaxies
 in less than 1 Gyr \citep{hibbard95,hibbard96}.

\subsection{\HI\ gas fractions}
Using the scaling relations from the $\alpha$40 catalog of the ALFALFA survey \citep{huang12}, 
star-forming galaxies with the stellar masses of J0836+30 and J0900+46 would be expected to 
have average \HI\ masses of approximately $1 \times 10^{10}$ M$_{\odot}$ and $8.7 \times 10^9$ M$_{\odot}$,
 yet we observe only $2 \times 10^8$ M$_{\odot}$ and $1.2 \times  10^9$ M$_{\odot}$, respectively. 
This implies that some fraction of the gas may have been ionized and heated and we are currently
only observing the remaining fraction of \HI. The external \HI\ clouds we observe represent the remnants of
 the original gas  reservoirs swept out of the host galaxy, and from the present data it is not
 clear whether the missing \HI\ was also expelled or heated or simply dissipated. Regardless, the
 displaced gas reservoir is required to complete the quenching of these galaxies and prevent further
star formation.

Extrapolating from the Kennicutt-Schmidt star formation law, the correlation 
between the \HI\ gas fraction and the star formation history traced by the 
NUV$-r$ color provides a rough indicator for the cold gas surface 
density \citep{huang12,zwaan13}.   Relative to gas-rich star-forming 
galaxies (mostly spirals) \citep[ALFALFA; ][]{huang12}, 
low-redshift transition galaxies with stellar masses greater than
 10$^{10}$ M$_{\odot}$ \citep[GASS; ][]{catinella12} and nearby 
early-type galaxies from the Atlas3D survey \citep{serra12,cappellari13,young13}, we
find that J1117-51 and J1237+36 have similar gas fractions and NUV$-r$ colors
 to E+A galaxies and straddle the region between star-forming galaxies 
with low gas fractions and that of gas-rich early-type galaxies.

\begin{figure}
\begin{center}
\includegraphics[scale=.68,angle=0]{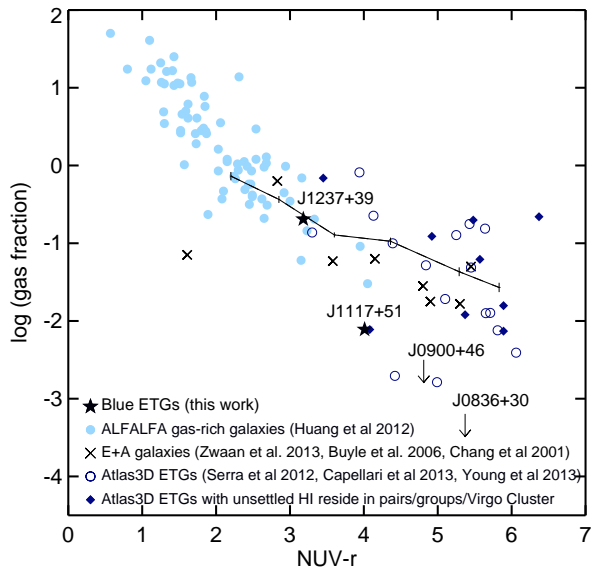}
\end{center}
\caption{\HI-to-stellar mass ratio as a function of galaxy color. The solid line
shows the average gas fractions found from GASS survey of massive transition-type galaxies \citep{catinella12}. }
\label{fhi}
\end{figure}

Figure~\ref{fhi} compares the \HI-stellar mass ratio and the integrated
 NUV$-r$ colour  of J1117+51 and J1237+39 (marked as black 
stars) to those of  E+A galaxies (represented by black crosses) where 
\HI\ has been previously detected \citep{chang01,buyle06,helmboldt07,zwaan13}.
galaxies from ALFALFA are represented by blue solid points and the solid line
 marks the average colours and average gas fractions found from GASS.
Near-ultraviolet (NUV) measurements of our sample
 were obtained from the Galex satellite telescope via the NASA Extragalactic 
Database\footnote{http://ned.ipac.caltech.edu}. 

We also compare our pilot sample of blue ETGs to recent \HI\ observations of
nearby ETGs from the Atlas3D survey \citep{serra12}.  The Atlas3D ETGs with 
regularly-rotating undisturbed \HI\ morphologies are represented by open circles, while,
Atlas3D galaxies with very disturbed and unsettled \HI\ morphologies and 
kinematics are represented by solid diamonds in Figure~\ref{fhi}.  It should 
be noted that all the Atlas3D ETGs with disturbed \HI\ morphologies reside in 
group or Virgo Cluster environments. Also, for a given NUV$-r$ colour,  these
Atlas3D galaxies have higher gas fractions than that of the E+A or blue ETG samples.
This suggests that either (1) the Atlas3D ETGs where \HI\ is observed are 
 at earlier stages of evolution than the sample of E+A galaxies or blue ETGs; or that (2) the merger scenario is as likely to result in an increase in gas fraction as it is to a reduction. 
Upper limits are shown for J0836+30 and J0900+46 due to the lack of \HI\
within the host galaxies.

We find the \HI\ gas fraction and NUV$-r$ colour of J1237+39 
(the galaxy at the earliest stage of quenching) to be very comparable to those of
star-forming galaxies from both the ALFALFA and GASS surveys.  On the other hand,
the gas fractions for the three more evolved blue ETGs in our sample are significantly lower
than the average gas fractions expected from star-forming galaxies, transitioning
galaxies or even gas-rich early-type galaxies with similar NUV$-r$ colours.
Similar to our results, the gas fractions of post-starburst E+A galaxies are lower than 
the average gas fractions for any given NUV$-r$ colours.   The combination of the \HI\ 
mapping and the radio continuum observations of  J0836+30 and J0900+46 
suggest that the star formation in these two galaxies will be truncated soon.
This result is consistent with those of \citet{schawinski09a} who were unable to 
detect any significant molecular gas reservoirs within Green Valley galaxies with Seyfert
 ionisation properties.


As previously seen in Figure~\ref{colcol}, these pilot \HI\ observations
show that the main mechanism for the fast quenching of star formation in blue
early-types is due to the physical displacement of the main gas reservoir from 
which stars are formed.   Hence, it is  likely that the depressed gas fractions
 from our sample (see Figure~\ref{fhi}) are due to a fast quenching process (relative
to other evolving galaxies) which removes a significant fraction, if not the entire gas reservoir,
 from which stars are formed.

Should the gas fall back, it may restart minor star formation which may be visible as the
frosting $\approx$ 1\% mass fraction of young stars often observed in quenched early-type galaxies
\citep{yi05,schawinski06,kaviraj07}.
While it may cause frosting, this returning gas however will not return the host galaxy to self-
regulated star-formation on the main sequence. An alternative fate for the ejected gas is that it
persists at large distances from the quenched galaxy, as is observed in the Milky Way \citep[in the form of the starless Magellanic stream; ][]{for14}; as well as in many quenched early-type
galaxies \citep[e.g.\ ][]{serra12}.

\subsection{Argument against coincident dark galaxies}
Dark galaxies can be typically defined as the extreme end of low surface 
brightness galaxies where few stars are found, consisting mainly of gas.  
With respect to our observed extraplanar gas clumps, it is very unlikely that 
these gas clumps are neighbouring dark galaxies.  Current re-ionization 
models of the Universe predict the latter as they find that the gas from 
95 per cent of the low-mass systems 
($M_{\rm{virial}} \leq 10^8$ M$_{\odot}$ or $v_{\rm{circ}} \leq 20$ km s$^{-1}$) 
appears to have been photoevaporated during the epoch of re-ionization \citep{susa04}.

Apart from a few reported cases of extragalactic gas clouds in group/cluster 
environments \citep{minchin05,davies06}) with no 
known optical counterparts, current and previous \HI\ all-sky surveys such
 as ALFALFA \citep{haynes11} and HIPASS \citep{wong06,doyle05} have not found any
 isolated extragalactic HI clouds devoid
 of stars.  This is consistent with simple gas equilibrium  models \citep{taylor05} 
which concluded that in the absence of an
 internal radiation field, dark galaxies/gas clouds with masses greater 
than $10^9$ M$_{\odot}$ will become Toomre unstable against star formation 
and start forming stars.
Galaxy evolution simulations  demonstrated that the majority of gas 
clouds identified with no known optical counterpart can be easily reproduced 
by galaxy--galaxy interactions \citep{bekki05,duc08}.

\section{Summary}
We have performed deep imaging of the \HI\ content and 1.4 GHz radio continuum emission 
of four blue early-type galaxies that are at four different stages of star formation truncation
using the WSRT. The \HI\ morphologies, kinematics and relative \HI-gas fractions are 
excellent probes and measures of the quenching evolutionary stages for each
of our sample galaxy.  A summary of our results are as follows:

\begin{enumerate}

\item We observe nuclear 1.4 GHz radio continuum emission that are consistent with emission from 
star formation from three galaxies at the three earliest stages of quenching evolution, namely,
 J1237+39, J1117+51 and J0900+46 (in the order of earliest to later stages of evolution).

\item The galaxy at the earliest stage of quenching, J1237+39, also has the bluest
NUV$-r$ colour, weakest [OIII/H$\beta$] ratio, the highest \HI\ gas fraction and a
symmetric rotating \HI\ disk.

\item At more advanced stages of quenching evolution, we observe increasingly asymmetric and 
increased spatial offsets between the \HI\ gas and the stellar component of the galaxy.   
Non-rotating \HI\ gas kinematics are also observed.  In the case of the two
galaxies at the most advanced stages of evolution where Seyfert ionisation signatures have been 
observed  (J0900+46 and J0836+30), the \HI\ gas reservoirs have been entirely
expelled by approximately 14--86 kpc from their respective host galaxies.  Due to the lack
of neighbouring galaxies, it is difficult to attribute the stripped gas to tidal or ram pressure interactions.

\item In the galaxy at the most advanced quenching stage, J0836+30, the expelled gas reservoir 
is  observed to be in alignment between the host galaxy and two radio lobes--- suggesting the
gas reservoir may have been swept out of the galaxy by powerful outflows from the central AGN 
in a previous active phase.  This scenario is consistent with recent observations of
AGN-driven gas outflows both locally and at higher redshifts \citep{sturm11,mahony13,morganti13,emonts14}.

\item  We conclude that the rapid quenching of star formation in low-redshift
early-type galaxies is due to the physical expulsion of the entire gas reservoir,
rather than the exhaustion of gas via star formation.  An AGN outflow
has the necessary energy to expel the reservoir of cold gas.
In addition, we do not think that AGN heating of the gas is a dominant mechanism
because the \HI\ gas would not remain visible as a coherent structure if this 
was the case.

\end{enumerate}

\vspace{1cm}

\chapter{\bf{Acknowledgments.}}
We thank the anonymous referee for their support of this project and for  improving the manuscript.  OIW acknowledges a Super Science Fellowship from the Australian Research Council
 and the Helena Kluyver visitor programme at ASTRON/JIVE.   KS is supported by SNF
 Grant PP00P2 138979/1. The Westerbork Synthesis Radio Telescope is operated by the
 ASTRON (Netherlands Institute for Radio Astronomy) with support from the Netherlands 
Foundation for Scientific Research (NWO). This publication makes use
of SDSS DR10 data.  This research has made use of the NASA/IPAC Extragalactic Database
(NED) which is operated by the Jet Propulsion Laboratory, California Institute of 
Technology, under contract with the National Aeronautics and Space Administration.

\bibliographystyle{mn2e} 
\bibliography{mn-jour,paperef}

\end{document}